\documentclass[prl,twocolumn,amsmath,amssymb]{revtex4}
\usepackage{amsmath}
\usepackage{graphicx, dsfont}
\usepackage{color}
\usepackage[breaklinks=false]{hyperref}

\begin{document}

 \begin{@twocolumnfalse}
 \begin{center}
    \mbox{
{
\color{blue} \href{http://link.aps.org/doi/10.1103/PhysRevLett.110.174103}{DOI: 10.1103/PhysRevLett.110.174103}
}
    }
\end{center}
  \end{@twocolumnfalse}

\title{Stochastic Hydrodynamic Synchronization in Rotating Energy Landscapes}
\author{N. Koumakis$^1$ and R. Di Leonardo$^{1,2}$}
\affiliation{
$^1$CNR-IPCF UOS Roma c/o Dip. Fisica, Universit\`a di Roma ``Sapienza", 00185 Rome, Italy\\
$^2$ Dipartimento di Fisica, Universit\`a di Roma ``Sapienza", 00185 Rome, Italy
}

\begin{abstract} 
Hydrodynamic synchronization provides a general mechanism for the spontaneous emergence of coherent beating states in independently driven mesoscopic oscillators. A complete physical picture of those phenomena is of definite importance to the understanding of biological cooperative motions of cilia and flagella. Moreover, it can potentially suggest novel routes to exploit synchronization in technological applications of soft matter.
We demonstrate that driving colloidal particles in rotating energy landscapes results in a strong tendency towards synchronization, favouring states where all beads rotate in phase. The resulting dynamics can be described in terms of activated jumps with transition rates that are strongly affected by hydrodynamics leading to an increased probability and lifetime of the synchronous states. Using holographic optical tweezers we quantitatively verify our predictions in a variety of spatial configurations of rotors.
\end{abstract}

\date{}
\maketitle

In the low Reynolds number world where cells and micro-organisms live, hydrodynamic interactions are so pervasive that dynamics are practically always a manybody problem \cite{happel}. In thermal equilibrium, this only affects time dependencies leaving static correlations uniquely defined by the energy function. In other words, a collection of still microscopy images will show signs of hydrodynamic interactions only when examined as a time sequence.  The situation can be dramatically different in off-equilibrium systems where hydrodynamics can induce striking static correlations, as in the coordinated motions of cilia and flagella that are often observed in living cells \cite{bray}. There is a growing consensus on the mainly hydrodynamic origin of those correlations \cite{raminrev, lauga}, although different mechanisms may be at work \cite{nohydrosync} and a final complete picture is still missing. The development of minimal physical models that display synchronization can provide new insights into the complex beating patterns found in nature, and at the same time finds technological importance in the design of novel strategies that exploit self-ordering of independently driven systems \cite{yan}. Minimal models usually assume as the basic oscillating/rotating unit a spherical bead  which is driven in periodic motion by a non-conservative force field. The phases of these basic units are not constrained by the applied forces, but are free to change in response to hydrodynamic couplings with nearby particles. This requirement of a free phase can be obtained in linear oscillators using a geometric switch, that is an external potential that switches between two shapes whenever some geometric trigger is activated based on particle instantaneous positions \cite{lagomarsino, kotar}. When a nonlinear potential is used, the obtained oscillations are asymmetric under time reversal and therefore can give rise to a quick synchronization of nearby oscillators \cite{bruot}. Although a mechanism for a geometric switch could be present in ciliar dynamics \cite{gueron}, it's implementation in an artificial oscillator requires a constant tracking feedback that makes it impractical for applications. Free phase periodic motions are more easily obtained by driving beads over closed two dimensional trajectories with prescribed force fields. Again kinematic reversibility prevents synchronization in the most simple case of strictly rigid trajectories and constant tangential forces \cite{lenz}. However, it has been found that the introduction of some mechanical flexibility can lead to metachronal waves \cite{brumley} or synchronization \cite{stark, caos}, although with strict requirements on the precise matching of applied torques \cite{minimal, rdl}. Orbital compliance is not a necessary condition and synchronization can additionally occur when velocity dependent forces are applied along elliptical tilted orbits near a wall \cite{vilfan}. Another route to synchronization consists of moving along strictly prescribed trajectories with purely tangential forces with a  phase profile that satisfies some generic conditions \cite{ramin}. This last case corresponds to the superposition of a static periodic potential to a constant tangential force field. Again the need for a non-conservative tangential force component makes practical realization of such rotors quite involved due to the constant need for a feedback mechanism unless a direct way of applying an external torque is available \cite{rdl}. An alternative way of inducing steady currents of microscopic objects is that of using time periodic sequences of spatially periodic conservative energy landscapes. These multi-state ratchets have been shown to be able to transport colloidal particles even in the presence of strong thermal noise and may have analogues in the biological realm \cite{koss}. 

In this Letter we propose a novel mechanism for hydrodynamic synchronization that is based on driving rotors with a periodic, conservative force field that is set into rotation. We show that the stability of the local minima in this potential is strongly affected by hydrodynamic interactions, giving the longest lifetimes and highest probabilities to synchronized states. Using dynamic holographic optical tweezers, we provide evidence for this phenomenon by the observation of colloidal beads in rotating optical landscapes. The observed dynamics can be described in terms of activated jumps between different phase configurations. We provide analytical expressions for the transition rates that quantitatively explain the robustness of this new mechanism.
\begin{figure}[t]
\includegraphics[width=.45\textwidth]{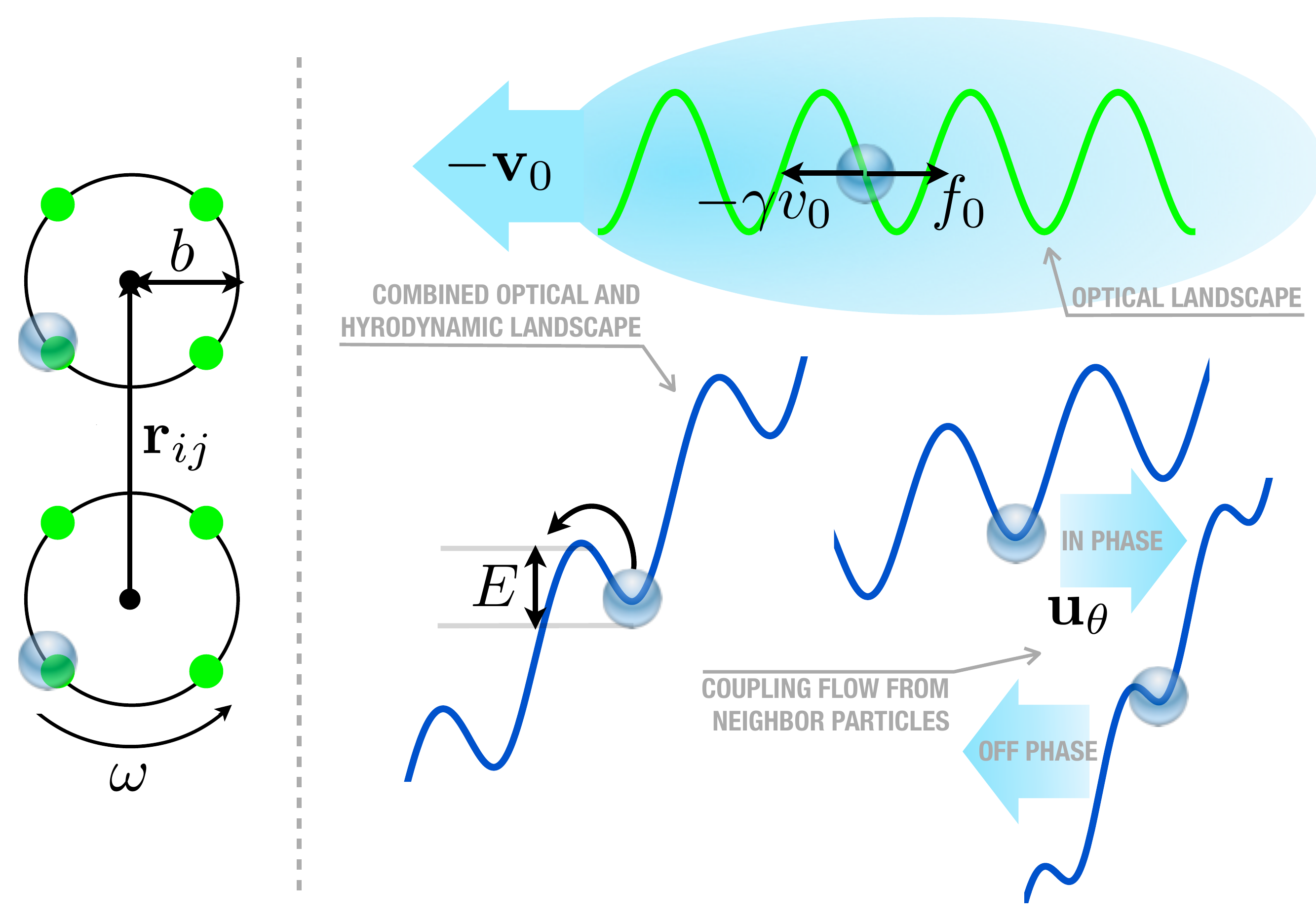}
\caption{Left panel: top view of a system composed by two beads trapped in rotating rings of optical traps (green spots). Right panel: schematic representation of  the optical landscape along the ring  (green line) and of the effective energy landscapes  that a bead experiences in the rotating frame (blue lines). Hydrodynamic drag due to background flow tilts the optical landscape by an amount that depends on the relative phase of a neighboring particle.}
\label{landscheme}
\end{figure}

In our approach, a colloidal particle is driven in periodic motion around a circular trajectory by rotating, with angular frequency $\omega$, an energy landscape composed by $k$ minima arranged on a ring of radius $b$. If we restrict our description to the azimuthal $\theta$ coordinate of the bead, we end up in the one dimensional problem of an overdamped particle over a periodic potential which travels with speed $\omega b$ (Fig. \ref{landscheme}). Moving to the rotating frame where the optical landscape is stationary, we need to consider the extra constant force due to the viscous drag exerted by the background flow with speed $-\omega b$. Due to this constant force the total energy landscape will be tilted, shifting the bead's equilibrum positions backward and lowering the energy barrier $E$ that a particle has to overcome for escape (Fig. \ref{landscheme}). If the rotating speed is large enough, the bead will randomly overcome that energy barrier with a probability per unit time given by Kramers formula \cite{risken}:

\begin{equation}
\lambda_0\propto \exp[-E/K_B T]
\end{equation}
Such a system constitutes a self-sustained rotator with a free phase and will be our basic synchronizing unit.  When a rotator $j$ is nearby to rotator $i$, it will contribute to the background flow experienced by $i$ with a tangential flow component given by: 
\begin{equation}
u_\theta(t)=\hat{\boldsymbol \theta}_i(t)\cdot G(\mathbf r_{ij})\cdot \hat{\boldsymbol  \theta}_j(t)\,f_0
\end{equation}
where $\hat{\boldsymbol \theta}_i=\{-\sin(\theta_i), \cos(\theta_i)\}$ is the tangential velocity versor of particle $i$ and $G(\mathbf r_{ij})=(8\pi\mu\, r_{ij})^{-1}(\mathds 1+\hat{\mathbf r}_{ij} \hat{\mathbf r}_{ij})$ is the Oseen tensor \cite{happel}. To simplify calculations, the Oseen tensor relating particle $i$ and $j$ is always evaluated for the constant position vector $\mathbf r_{ij}$ joining the centers of the two rings as shown in Fig. \ref{landscheme}.  Moreover we neglect fluctuations in the tangential force driving the particles in their rotary motions and assume it has the constant value $f_0$. With these approximations it is straightforward to calculate the total background flow on particle $i$:
\begin{equation}
\label{u}
u_\theta(t)=\frac{f_0}{8\pi\mu}\sum_{j\neq i}\frac{1}{r_{ij}}
\left[\cos(\theta_i-\theta_j)+\sin\theta_i\sin\theta_j\right]
\end{equation}
that depends on time through $\theta_i=\varphi_i+\omega t$. The phase variable $\varphi_i$ hops between the discrete values $2\pi k/K$, with $K$ the number of traps.  Assuming the hopping rate of $\varphi_i$ is small compared to $\omega$, we can average out the time dependence in (\ref{u}) by integrating over one cycle with frozen $\varphi_i$. The time dependence cancels out in the first term while the second term $\sin\theta_i\sin\theta_j$ will oscillate at a frequency $2\omega$ around the mean value $\cos(\varphi_i-\varphi_j)/2$. If we now call $\ell$ the length separating the stable equilibrium point in a running trap from the energy barrier maximum we find that nearby particles provide additional drag forces that, at the lowest order, affect the average energy barrier by:
\begin{equation}
\overline{\delta E_i}=\frac{9 \gamma v_0 \ell}{8}\sum_{j\neq i}\frac{a}{r_{ij}}\cos(\varphi_i-\varphi_j)
\label{Ei}
\end{equation}
where $\gamma=6\pi\eta a$ is the Stokes drag coefficient depending on the viscosity $\eta$ and particle's radius $a$. Neglecting fluctuations around this average value, the escape rate for particle $i$ retains the Kramers formula evaluated for the time average of the fluctuating barrier:
\begin{equation}
\lambda_i=\lambda_0 \exp\left[-\overline{\delta E_i}/K_B T\right]
\label{li}
\end{equation}
Combining equations (\ref{Ei}) and (\ref{li}) we realize that in-phase neighbors ($\varphi_j=\varphi_i$) will contribute to stabilize the phase of particle $i$ by increasing the average energy barrier for escape. On the contrary off-phase neighbors ($\varphi_j=\varphi_i+\pi$) will increase the escape rate.
Previous investigations of minimal models have evidenced that the phase difference between two hydrodynamically coupled rotors evolves like a Brownian particle over a tilted periodic potential whose minima represent synchronized states \cite{minimal, goldstein, rdl}. The amount of tilt is due to the mismatch in the natural frequencies of the two oscillators while the depth of the minima measures the effectiveness of the proposed mechanism in favoring synchronized states.
On the other hand, the tilted washboard potential in Fig. \ref{landscheme} represents the combined optical and hydrodynamic landscape that is experienced by an isolated bead in its own rotating frame. Each rotator is now represented by a phase variable that evolves through a discrete set of allowed values by means of activated jumps. A combined system of rotators will be described by a phase vector whose components evolve by discrete jumps with rates defined by (\ref{li}). Our prediction is that the resulting master equation will display a stationary state with an increased probability for synchronized configurations. This is an intrinsically stochastic model, that has no deterministic analogue and where none of the previously recognized ingredients is playing a role: there's no radial compliance in our model, no anisotropy or wall effects. A phase dependent force may suggest an analogy with the static phase dependent force model in \cite{ramin} but that only works when specific conditions are met. On the contrary, all arguments used to derive equations (\ref{Ei}) and (\ref{li}) remain valid for a multistable potential of a generic shape.

Experimentally we have implemented such a system through rotating optical landscapes produced using holographic tweezers \cite{grier, hots}. Our setup uses a reflective liquid crystal space light modulator (Holoeye LC-2500) to impose a computer generated pattern of phase shifts \cite{gsw} on an expanded laser beam (DPSS Opus Ventus 532 nm, 3W). The emerging wavefront is focused by a microscope objective (Nikon Plan Apo VC 100x, NA 1.4) onto an array of trapping spots arranged on two or more nearby rings (Fig \ref{landscheme}). Particle rotation is achieved by rapidly rotating traps with a cycle of  three holograms \cite{koss}. 

\begin{figure}[t]
\includegraphics[width=.45\textwidth]{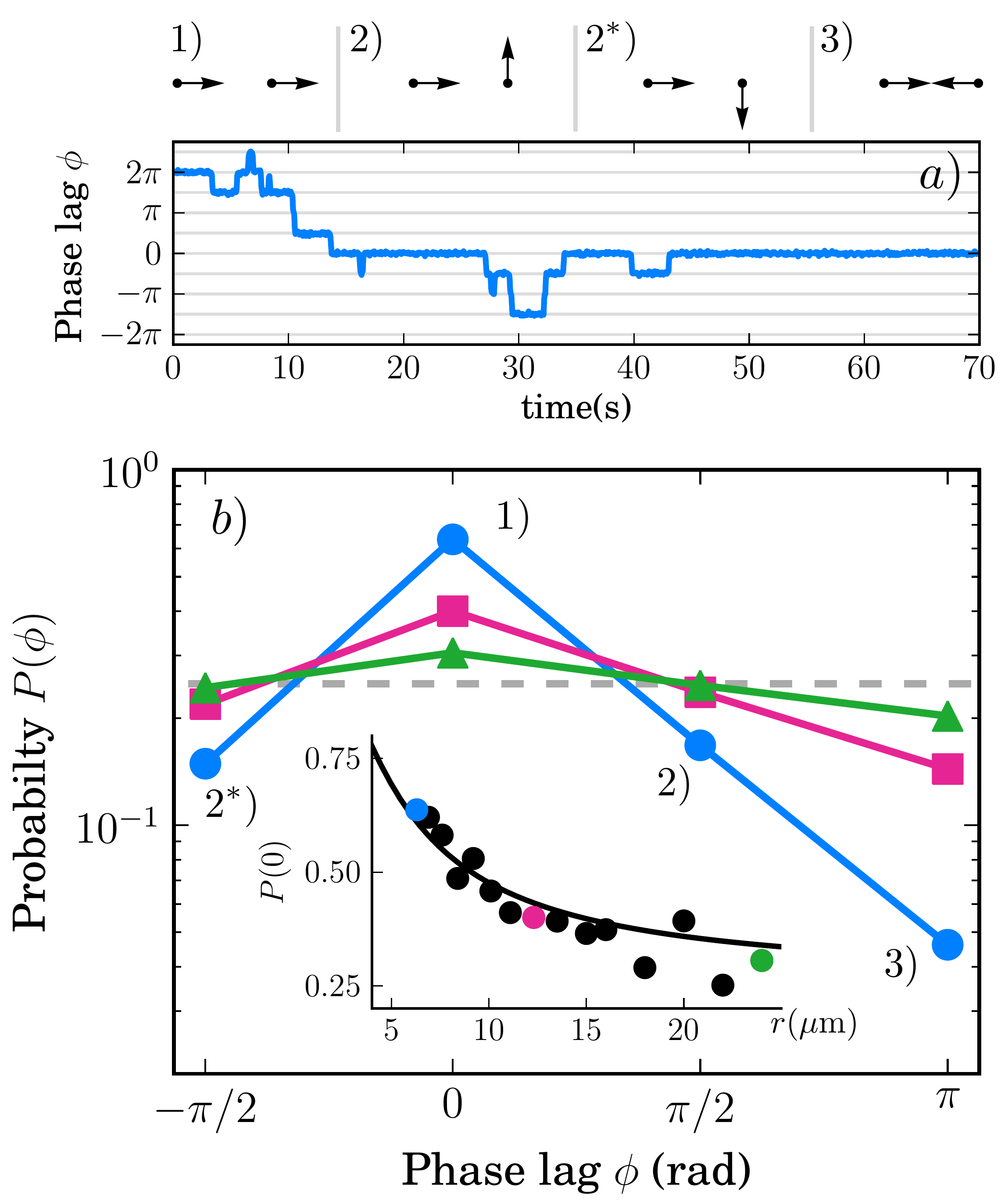}
\caption{
a)Time evolution of phase lag $\phi$. b)Probability distribution of the phase lag between two adjacent rotors. Center to center distances are 6.3, 12 and 24 $\mu$m for respectively circles, squares and triangles.  Dashed line represents the uncoupled limit of 1/4. The inset shows the probability of the syncronised state ($\phi=0$) as a function of distance between the rings together with a fit to the theory described in the text (solid line).}
\label{2ring}
\end{figure}

\begin{figure}[t]
\includegraphics[width=.45\textwidth]{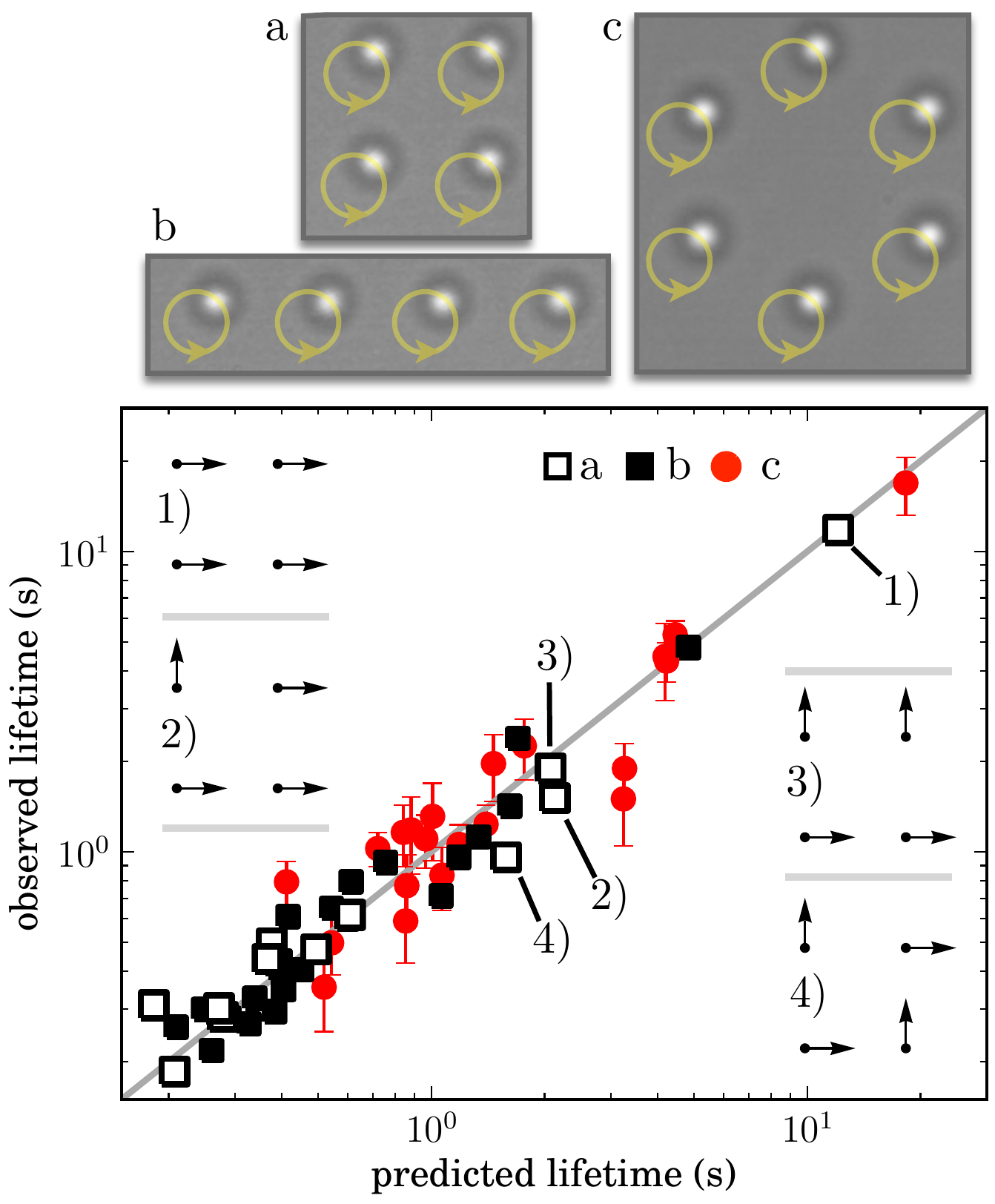}
\caption{Observed average lifetimes of phase configurations vs theoretically predicted lifetimes for different rotor arrangements as shown in the panels on top. b) open squares:  four rotors on a square, c) solid squares: four rotors on a line, d) solid circles: six rotors on a ring. The phases of the four longest lived configurations for geometry b) are represented as arrows.}
\label{ltplot}
\end{figure}

In the simple two particle case, different configurations are defined through the single phase variable $\phi=\theta_i-\theta_j$. According to (\ref{li}), a generic state $\phi$ will have a lifetime:
\begin{equation}
\tau(\phi)={\tau_\infty}\exp\left[\alpha \frac{a}{r}\cos\phi\right]
\label{2ringlt}
\end{equation}
where $\tau_\infty=(2\lambda_0)^{-1}$ is the lifetime corresponding to infinite separation and the factor of 2 accounts for the fact that a generic configuration can change when either of the two particles suffers a phase slip. 
Hydrodynamic interactions affect transition rates exponentially and if the adimensional constant $\alpha=9 \gamma v_0 \ell/8 K_B T$ is large enough, synchronized states ($\phi=0$) should acquire a much longer lifetime. In Fig. \ref{2ring}a) we report a sample time trace of the phase lag between two nearby rotors at a distance of 6.3 $\mu$m. In all of the experiments reported here each particle moves on a ring of four traps of radius $b=$1.5 $\mu$m that rotates at a frequency of $1.7$ Hz. The relative phase between two particles can only assume values given by an integer multiple of $\pi/2$. The trace clearly shows that the system evolves by sudden $\pi/2$ jumps that become remarkably less frequent as we approach the synchronous state $\phi=0$ (see supplementary movie). Fig. \ref{2ring}b) shows the histograms of observed phase lags $\phi$ as obtained by averaging over several traces spanning a total time of about $2000$ seconds. There is a total of four different states corresponding to the phase lags $\{-\pi/2,0,\pi/2,\pi\}$ and whose lifetimes can be anticipated using Eq.(\ref{2ringlt}). Having an expression for all transition rates, we can write the corresponding master equation for the occupation probabilities of the four states. For two particles this can be done analytically and gives:
\begin{equation}
\label{p2}
P(\phi)=\tau(\phi)/\sum_\phi \tau(\phi)
\end{equation}
In particular it predicts that the probability of states $\{0,\pi/2,\pi\}$ should fall on a straight line of slope $\alpha a/r$ once plotted on a logarithmic scale, which is indeed observed in Fig. \ref{2ring}. At the closest ring separation, the synchronized state with $\phi=0$ occurs with a probability $P(0)=0.63$  which is 14 times higher than the anti-phase state $P(\pi)=0.045$. Through the equivalent ratio of lifetimes, as taken from equation (\ref{2ringlt}) we find an $\alpha=8.2\pm$0.3, from which we can extract $\ell=0.1$ $\mu$m. By increasing the distance $r$ between the rings, we find that $P(0)$ decays to the uncoupled limit of 0.25 (inset of Fig. \ref{2ring}), while fitting the data with Eq.(\ref{p2}) produces an $\alpha=7.9\pm$0.4, which confirms the hydrodynamic origin of the observed synchronization. 
Through (\ref{2ringlt}) we find that when particle size and speed are fixed, the stability of the synchronized state can be enhanced by increasing $\ell$. This can be experimentally achieved either by using stronger traps or by increasing distance between them.
For a number of rotors $N$ larger than two we can obtain the escape rate from a given configuration of relative phases as the sum of the escape rates for each rotor and get the lifetime as the inverse of the resulting rate:
\begin{equation}
\label{lt}
\tau(\{\theta_i\})=\tau_\infty
\left(
\frac{1}{N}
\sum_{i} \exp\left[-\overline{\delta E_i}/K_BT\right]
\right)^{-1}
\end{equation}
where now $\tau_\infty=(N\lambda_0)^{-1}$. The former expression allows to easily calculate  the lifetime of any state as a function of directly measurable quantities with the only exception of the length $\ell$. To test Eq.(\ref{lt}) we have investigated three different 2D geometries: four rotors on a square, four rotors on a line and six rotors on a ring. In all these different geometries we kept the nearest neighbor distance fixed to 6.3 $\mu$m. Fig.\ref{ltplot} reports the average lifetimes measured from experimental trajectories plotted against their theoretical expected values according to Eq. (\ref{lt}). For all configurations we find that the longest lived state is the one with all particles rotating in phase, a condition required to maximize all $\overline{\delta E_i}$. By using the length $\ell$ as a fitting parameter, we find the agreement is very good, providing support to all the approximations used for the lifetime derivation. In particular, we find that the effect of hydrodynamic couplings between particles is very sensitive to the actual phase configuration, thus giving rise to a broad spectrum of lifetimes which spans almost two decades. When keeping the number of particles fixed, the lifetime of the synchronized state becomes larger for more compact geometries as evidenced by comparing the data relative to four beads arranged on a square and on a line. Similarly, if we keep the nearest neighbor distances constant and add particles, as examined with four and six particles (a) and c) in Fig. \ref{ltplot}), we find that the lifetime of the synchronized state is increased. 

It is worth noting that upon increasing the number of particles $N$ the corresponding number of possible phase configurations increases exponentially as $k^{N-1}$ with $k$ the number of minima or traps. In the absence of hydrodynamic interactions the probability of a specific state such as the synchronous state would rapidly decay to zero. In particular,  we would have a probability of $4^{-3}$=0.016 for $N$=4 and $4^{-5}$=0.001 for $N$=6.  However hydrodynamic interactions are so effective in favoring synchronous states that we observe them occurring with probabilities of 0.4  and 0.2 for configurations a) and c) in Fig. \ref{ltplot}. Synchronous states are favored by a combination of a long lifetime and the form of individual particle escape rates as described in Eq. \ref{li}. In particular, the transition rates are higher for those particles that rotate with the largest phase deviations leading to a driving force towards synchronization.  

Previous studies have addressed the connection of synchronized states with states of extremal dissipation (minimal or maximal) \cite{raminrev}. It is interesting to note that, as opposed to other minimal models of rotating beads, the synchronous states in our model always display minimal dissipation. In general, synchronous states correspond to minimal viscous drag.  When the beads are driven by constant forces this results in higher speeds and therefore higher dissipation. In our case the beads move in tandem with an energy landscape which is rotated at a constant speed, that means that a smaller driving force is needed for in phase motions, leading to the minimization of expended energy.

In conclusion, we have demonstrated that hydrodynamic interactions are capable of inducing strong static correlations among particles that are driven by rotating multistable potentials. Those interactions always lead to an exponential increase in the lifetime of synchronous states, where all particles rotate in phase. Moreover, when increasing the number of particle $N$, we find that the probability of observing full synchronization remains large even if the total number of states has grown exponentially with $N$. This novel mechanism is intrinsically stochastic and requires no special condition other than driving particles with a rotating landscape. Although mechanical models based on orbit compliance seem to be very successful in describing eukaryotic flagella \cite{regrowth}, modeling dynamics through activated jumps might be more suited for the description of bacterial flagella, which are known to rotate in discrete steps \cite{sowa}. It will also be interesting to extend our studies to the case of particle transport along linear landscapes with possible connections to intracellular transport along cytoskeletal filaments \cite{bray}.

We acknowledge funding from IIT-SEED BACTMOBIL project and MIUR-FIRB project
No. RBFR08WDBE.  RDL also acknowledges support from ERC Staring Grant SMART.
\bibliographystyle{apsrev}

\begin{thebibliography}{22}

\bibitem{happel}
J. Happel and H. Brenner, {\it Low Reynolds Number Hydrodynamics} (Kluwer Academic, Dordrecht, 1983).

\bibitem{bray}
D. Bray, {\it Cell Movements: From Molecules to Motility} (Garland Publishing, New York, 2000).

\bibitem{raminrev}
R. Golestanian, J. M. Yeomans, and N. Uchida,
Soft Matter {\bf 7}, 3074-3082 (2011).

\bibitem{lauga}
E. Lauga and T. R. Powers,
Reports on Progress in Physics {\bf 72}, 096601 (2009).

\bibitem{nohydrosync}
B. M. Friedrich and F. J\"ulicher,
Phys. Rev. Lett. {\bf 109}, 138102 (2012).

\bibitem{yan}
J. Yan, M. Bloom, S. C. Bae, E. Luijten, and S. Granick,
Nature {\bf 491}, 578--581 (2012).

\bibitem{lagomarsino}
M. C. Lagomarsino, P. Jona, and B. Bassetti,
Phys. Rev. E {\bf 68}, 021908 (2003).

\bibitem{kotar}
J. Kotar, M. Leoni, B. Bassetti, M. C. Lagomarsino, and P. Cicuta,
Proc. Natl. Acad. Sci. {\bf107}, 7669-7673 (2010).

\bibitem{bruot}
N. Bruot, J. Kotar, F. de Lillo, M. Cosentino Lagomarsino, and P. Cicuta,
Phys. Rev. Lett. {\bf 109}, 164103 (2012).

\bibitem{gueron}
S. Gueron and K. Levit-Gurevich,
Proceedings of the National Academy of Sciences {\bf 96}, 12240-12245 (1999).

\bibitem{lenz}
P. Lenz and A. Ryskin,
Physical Biology {\bf 3}, 285 (2006).

\bibitem{brumley}
D. R. Brumley, M. Polin, T. J. Pedley, and R. E. Goldstein,
Phys. Rev. Lett. {\bf 109}, 268102 (2012).

\bibitem{stark}
M. Reichert and H. Stark,
The European Physical Journal E {\bf17}, 493-500 (2005).

\bibitem{caos}
T. Niedermayer, B. Eckhardt, and P. Lenz,
Chaos {\bf18}, 037128 (2008).

\bibitem{minimal}
B. Qian et al. Phys. Rev. E {\bf 80}, 061919 (2009).

\bibitem{rdl}
R. Di Leonardo, A. B\'uz\'as, L. Kelemen, G. Vizsnyiczai, L. Oroszi, and P. Ormos,
Phys. Rev. Lett. {\bf 109}, 034104 (2012).

\bibitem{vilfan}
A. Vilfan and F. J\"ulicher,
Phys. Rev. Lett. {\bf 96}, 058102 (2006).

\bibitem{ramin}
N. Uchida and R. Golestanian,
Phys. Rev. Lett. {\bf 106}, 058104 (2011).

\bibitem{koss}
B. A. Koss and D. G. Grier,
Applied physics letters {\bf 82}, 3985--3987 (2003).

\bibitem{risken}
H. Risken, The Fokker-Planck equation, Springer-Verlag, Berlin (1989)

\bibitem{goldstein}
R. E. Goldstein, M. Polin, and I. Tuval,
Phys. Rev. Lett. {\bf103}, 168103 (2009).

\bibitem{grier}
D. G. Grier,
Nature {\bf 424}, 810--816 (2003).

\bibitem{hots}
M. J. Padgett and R. D. Leonardo,
Lab Chip {\bf 11}, 1196-1205 (2011).

\bibitem{gsw}
R. Di Leonardo, F. Ianni, and G. Ruocco,
Optics Express {\bf 15}, 1913 (2007).

\bibitem{regrowth}
R. E. Goldstein, M. Polin, and I. Tuval,
Phys. Rev. Lett. {\bf 107}, 148103 (2011).

\bibitem{sowa}
Y. Sowa, A. D. Rowe, M. C. Leake, T. Yakushi, M. Homma, A. Ishijima, and R. M. Berry,
Nature {\bf 437}, 916--919 (2005).



\end{thebibliography}

\end{document}